\documentclass[final]{ols}

\usepackage{url}
\usepackage{zrl}

\ifpdf
\usepackage[pdftex]{graphicx}
\else
\usepackage{graphicx}
\fi

\begin{document}


\title{Optimizing real-world applications with GCC Link Time Optimization}
\subtitle{ }  
\date{}             

\shortauthor{Taras Glek, Jan Hubi\v cka}  

\author{
Taras Glek\\
{\itshape Mozilla Corporation}\\
{\ttfamily\normalsize tglek@mozilla.com}\\
\and
Jan Hubi\v cka\\
{\itshape SuSE \v CR s.r.o.}\\
{\ttfamily\normalsize jh@suse.cz}\\
} 

\maketitle

\begin{abstract}
GCC has a new infrastructure to support a link time optimization (LTO). The
infrastructure is designed to allow linking of large applications
using a special mode (WHOPR) which support parallelization of the compilation process.
In this paper we present overview of the design and
implementation of WHOPR and present test results of its behavior when optimizing large applications. We
give numbers on compile time, memory usage and code quality comparisons to the
classical file by file based optimization model. In particular we focus on Firefox web
browser. We show main problems seen only when compiling a large
application, such as startup time and code size growth.

\end{abstract}

\section{Introduction}
{\em Link Time Optimization (LTO)} is a compilation mode in which an intermediate
language (an {\em IL}) is written to the object files and the optimizer is invoked
during the linking stage. This allows the compiler to extend the scope of
inter-procedural analysis and optimization to encompass the whole program
visible at link-time. This gives the compiler more freedom than the file-by-file
compilation mode, in which each compilation unit is optimized independently,
without any knowledge of the rest of the program being constructed.

Development of the LTO infrastructure in the GNU Compiler Collection ({\em
GCC}) started in 2005 \cite{LTOproposal} and the initial implementation was
first included in GCC 4.5.0, released in 2009.  The inter-procedural
optimization framework was independently developed starting in 2003
\cite{callgraph}, and was designed to be used both independently and in tandem
with LTO.

The LTO infrastructure represents an important change to the compiler, as well
as the whole tool-chain. It consists of the following components:
\begin{enumerate}
  \item A middle-end (the part of GCC back-end independent of target
architecture) extension that supports streaming an intermediate
language representing the program to disk,
  \item A new compiler front-end (the LTO front-end), which is able to read back the intermediate
language, merge multiple units together, and process them in the compiler's
optimizer and code generation backend,
  \item A linker plugin integrated into the Gold linker, which is able to call back
into the LTO front-end during linking \cite{plugin}, 

(The plugin interface is designed to be independent of both the Gold linker and the
rest of GCC's LTO infrastructure; thus the effort to the extend the tool-chain for plugin
support can be shared with other compilers with LTO support. Currently
it is used also by LLVM \cite{LLVM}.)
  \item Modifications to the GCC driver (\code{collect2}) to support linking of
LTO object files using either the linker plugin or direct invocation of the
LTO front-end,
  \item Various infrastructure updates, including a new symbol table representation and support 
for merging of declarations and types within the middle-end, and
  \item Support for using the linker plugin for other components of the
tool-chain---such as {\tt ar} and {\tt nm}.  (Libtool was also updated to
support LTO.)
\end{enumerate}
The inter-procedural optimization infrastructure consists of the following major components:
\begin{enumerate}
  \item Callgraph and varpool data structures representing the program in
optimizer friendly form,
  \item Inter-procedural dataflow support,
  \item A pass manager capable of executing inter-procedural and local passes, and
  \item A number of different inter-procedural optimization passes.
\end{enumerate}
Sections~\ref{ltosect} and \ref{ipasect} contains an overview with more details
on the most essential components of both infrastructures.

GCC is the third free software C/C++ compiler with LTO support (LLVM and Open64
both supported LTO in their initial respective public releases).  GCC~4.5.0's
LTO support was sufficient to compile small- to medium-sized C, C++ and Fortran
programs, but had were several deficiencies, including incompatibilities with
various language extensions, issues mixing multiple languages, and the
inability to output debug information.  In Section~\ref{largesect} we describe
an ongoing effort to make LTO useful for large real-world applications, discuss
existing problems and present early benchmarks.  We focus on two applications
as  a running example thorough the paper: the GCC compiler itself and the
Mozilla Firefox browser.


\section{Design and implementation of the Link Time Optimization in GCC}
\label{ltosect}
The link-time optimization in GCC is implemented by storing the intermediate
language into object files.  Instead of producing ``fake'' object files of
custom format, GCC produces standard object files in the target format (such as
ELF) with extra sections containing the intermediate language, which is used
for LTO.  This ``fat'' object format makes it easier to integrate LTO into
existing build systems, as one can, for instance, produce archives of the
files.  Additionally, one might be able to ship one set of fat objects which
could be used both for development and the production of optimized builds,
although this isn't currently feasible, for reasons detailed below.  As a
surprising side-effect, any mistake in the tool chain that leads to the LTO
code generation not being used (e.g. an older libtool calling ld directly)
leads to the silent skipping of LTO.  This is both an advantage, as the system
is more robust, and a disadvantage, as the user isn't informed that the
optimization has been disabled.

The current implementation is limited in that it only produces ``fat'' objects,
effectively doubling compilation time.  This hides the problem that some tools,
such as {\tt ar} and {\tt nm}, need to understand symbol tables of LTO sections.
These tools were extended to use the plugin infrastructure, and with these
problems solved, GCC will also support ``slim'' objects consisting of the
intermediate code alone.

The GCC intermediate code is stored in several sections:
\begin{itemize}
 \item {\bf Command line options (\code{.gnu.lto_.opts})}

 This section
contains the command line options used to generate the object files.  This is
used at link-time to determine the optimization level and other settings when
they are not explicitly specified at the linker command line.

At the time of writing the paper, GCC  does not support combining LTO
object files compiled with different set of the command line options into
a single binary.

\item {\bf The symbol table (\code{.gnu.lto_.symtab})}

This table replaces
the ELF symbol table for functions and variables represented in the LTO IL.
Symbols used and exported by the optimized assembly code of ``fat'' objects
might not match the ones used and exported by the intermediate code.  The
intermediate code is less-optimized and thus requires a separate symbol table.

There is also possibility that the binary code in the ``fat'' object will lack a call to a
function, since the call was optimized out at compilation time after the
intermediate language was streamed out.  In some special cases, the same optimization
may not happen  during the link-time optimization. This would lead to an
undefined symbol if only one symbol table was used.

 \item {\bf Global declarations and types (\code{.gnu.lto_.decls}).}

This section contains an intermediate language dump of all declarations and
types required to represent the callgraph, static variables and top-level debug
info.
 \item {\bf The callgraph (\code{.gnu.lto_.cgraph}).}

This section contains
the basic data structure used by the GCC inter-procedural optimization
infrastructure (see Section~\ref{ipapass}). This section stores an
annotated multi-graph which represents the functions and call sites as well as the variables, aliases
and top-level {\tt asm} statements.
 \item {\bf IPA references (\code{.gnu.lto_.refs}).}

This section contains references
between function and static variables.
 \item {\bf Function bodies}

This section contains function bodies in the
intermediate language representation. Every function body is in a separate
section to allow copying of the section independently to different object files
or reading the function on demand.
 \item {\bf Static variable initializers\\ (\code{.gnu.lto_.vars}).}
 \item {\bf Summaries and optimization summaries used by IPA passes.}

See Section~\ref{ipapass}.
\end{itemize}

The intermediate language (IL) is the on-disk representation of GCC GIMPLE \cite
{GIMPLE}. It is used for high level optimization in the SSA form \cite{SSA}. The
actual file formats for the individual sections are still in a relatively early stage
of development.  It is expected that in future releases the representation will be
re-engineered to be more stable and allow redistribution of object files containing LTO
sections. Stabilizing the intermediate language format will require
a more formal definition of the GIMPLE language itself.  This is mentioned as one of
main requirements in the original proposal for LTO \cite{LTOproposal}, yet
five years later we have to admit that work on this made almost no progress.

\subsection{Fast and Scalable Whole Program Optimizations --- WHOPR}

One of the main goals of the GCC link-time infrastructure was to allow
effective compilation of large programs.  For this reason GCC implements two
link-time compilation modes.
\begin{enumerate}
\item  {\em LTO mode}, in which the whole program is read into the compiler
at link-time and optimized in a similar way as if it were a single source-level compilation unit. 

\item {\em WHOPR\footnote{An acronym for ``scalable WHole PRogram optimizer'',
not to be confused with the {\tt -fwhole-program} concept described later.}
mode} which was designed to utilize multiple CPUs and/or a distributed
compilation environment to quickly link large applications\footnote{Distributed
compilation is not implemented yet, but since the parallelism is facilitated
via generating a {\tt Makefile}, it would be easy to implement.} \cite{WHOPR}.
\end{enumerate}

WHOPR employs three main stages:
\begin{enumerate}
\item {\bf Local generation ({\em LGEN})}

This stage executes in parallel. Every file in the program is compiled into the
intermediate language and packaged together with the local call-graph and summary
information.  This stage is the same for both the LTO and WHOPR compilation mode.
\item {\bf Whole Program Analysis ({\em WPA})}

WPA is performed sequentially. The global call-graph is generated, and a
global analysis procedure makes transformation decisions. The global call-graph
is partitioned to facilitate parallel optimization during phase 3. The results of the
WPA stage are stored into new object files which contain the partitions of program
expressed in the intermediate language and the optimization decisions.
\item {\bf Local transformations ({\em LTRANS})}

This stage executes in parallel. All the decisions made during phase 2 are
implemented locally in each partitioned object file, and the final object code is generated.
Optimizations which cannot be decided efficiently during the phase 2 may be
performed on the local call-graph partitions.
\end{enumerate}
WHOPR can be seen as an extension of the usual LTO mode of compilation. In LTO,
WPA and LTRANS and are executed within an single execution of the
compiler, after the whole program has been read into  memory.

When compiling in WHOPR mode the callgraph partitioning is done during the WPA
stage.  The whole program is split into a given number of partitions of about
same size, with the compiler attempting to minimize the number of references
which cross partition boundaries.  The main advantage of WHOPR is to allow the
parallel execution of LTRANS stages, which are the most time-consuming part of
the compilation process. Additionally, it avoids the need to load the whole
program into memory.

The WHOPR compilation mode is broken in GCC~4.5.x.  GCC~4.6.0 will be the first
release with a usable WHOPR implementation, which will by default replace the
LTO mode.  In this paper we concentrate on the real-world behavior of WHOPR.

\subsection{Inter-procedural optimization infrastructure}
\label{ipapass}

The program is represented in a {\em callgraph} (a multi-graph
where nodes are functions and edges are call sites) and the {\em varpool} (a list of
static and external variables in the program) \cite{callgraph}.

The inter-procedural optimization is organized as a sequence of individual passes, which
operate on the callgraph and the varpool.  To make the implementation of WHOPR possible,
every inter-procedural optimization pass is split into several stages that are executed at
different times of WHOPR compilation:
\begin{itemize}
\item LGEN time:
\begin{itemize}
\item[1.] {\bf Generate summary}

Every function body and variable initializer is examined
and the relevant information is stored into a pass-local data structure.
\item[2.] {\bf Write summary}

Pass-specific information is written into an object
file.
\end{itemize}
\item WPA time:
\begin{itemize}
\item[3.] {\bf Read summary}

The pass-specific information is read back into
a pass-local data structure in memory.
\item[4.] {\bf Execute}

The pass performs the inter-procedural propagation. This
must be done without actual access to the individual function bodies or variable
initializers.  In the future we plan to implement functionality to bring
a function body into memory on demand, but this should be used with a care
to avoid memory usage problems.
\item[5.] {\bf Write optimization summary}

The result of the inter-procedural propagation
is stored into the object file.
\end{itemize}
\item LTRANS time:
\begin{itemize}
\item[6.] {\bf Read optimization summary}

Inter-procedural optimization decisions
are read from an object file.
\item[7.] {\bf Transform}

The actual function bodies and variable initializers are updated
based on the information passed down from the {\em Execute} stage.
\end{itemize}
\end{itemize}

The implementation of the inter-procedural passes are shared between LTO, WHOPR and
classic non-LTO compilation.
During the file-by-file mode
every pass executes its own {\em Generate summary}, {\em Execute}, and
{\em Transform} stages within the single execution context of the compiler. In LTO compilation mode every pass uses {\em Generate summary}, {\em Write
summary} at compilation time, while the {\em Read summary}, {\em Execute}, and {\em Transform} stages are executed at link time.
In WHOPR mode all stages are used.

One of the main challenges of introducing the WHOPR compilation mode was
solving interactions between the optimization passes.  In LTO
compilation mode, the passes are executed in a sequence, each of which consists of
analysis (or {\em Generate summary}), propagation (or {\em Execute}) and {\em Transform}
stages. Once the work of one pass is finished, the next pass sees the updated program
representation and can execute. This makes the individual passes
independent on each other.

In the WHOPR mode all passes first execute their {\em Generate summary} stage.
Then the summary writing ends LGEN.  At WPA time the summaries are read back
into memory and all passes run {\em Execute} stage.  Optimization summaries are
streamed and shipped to LTRANS. Finally all passes execute the {\em Transform}
stage.

Most optimization passes split naturally into analysis, propagation
and, transformation stages. The main problem arises when one pass performs
changes and the following pass gets confused by seeing different callgraphs at
the {\em Transform} stage than at the {\em Generate summary} or the {\em
Execute} stage.  This means that the passes are required to communicate
their decisions with each other. Introducing an interface in between each pair of optimization
passes would quickly make the compiler unmaintainable.

For this reason, the GCC callgraph infrastructure implements a method of
representing the changes performed by the optimization passes in the
callgraph without needing to update function bodies.

A {\em virtual clone} in the callgraph is a function that has no associated
body, just a description how to create its body based on a different function
(which itself may be a virtual clone) \cite{callgraph2007}. 

The description of function modifications includes adjustments to the function's
signature (which allows, for example, removing or adding function arguments),
substitutions to perform on the function body, and, for inlined functions, a pointer
to function it will be inlined into.

It is also possible to redirect any edge of the callgraph from a function to its
virtual clone.  This implies updating of the call site to adjust for the new function
signature.

Most of the transformations performed by inter-procedural optimizations can be
represented via virtual clones: For instance, a constant propagation pass can produce a virtual
clone of the function which replaces one of its arguments by a constant. The inliner can
represent its decisions by producing a clone of a function whose body will be
later integrated into given function.

Using virtual clones the program can be easily updated at the {\em Execute} stage,
solving most of pass interactions problems that would otherwise occur at
the {\em Transform} stages.  Virtual functions are later materialized in the
LTRANS stage and turned into real functions.  Passes executed after the virtual
clone were introduced also perform their {\em Transform} stages on new
functions, so for a pass there is no significant difference between operating
on a real function or a virtual clone introduced before its {\em Execute}
stage.

Optimization passes then work on virtual clones introduced before their {\em
Execute} stage as if they were real functions. The only difference is that
clones are not visible at {\em Generate Summary} stage.

To keep the function summaries updated, the callgraph interface allows an optimizer to
register a callback that is called every time a new clone is introduced as well
as when the actual function or variable is generated or when a function or
variable is removed.  These hooks are registered at the {\em Generate summary}
stage and allow the pass to keep its information intact until the {\em
Execute} stage. The same hooks can also be registered at the {\em Execute}
stage to keep the optimization summaries updated for the {\em Transform} stage.

To simplify the task of generating summaries several data structures in
addition to the callgraph are constructed.  These are used by several passes.

We represent IPA references in the callgraph.  For a function or variable
$A$, the {\em IPA reference} is a list of all locations where the address of $A$ is
taken and, when $A$ is a variable, a list of all direct stores and reads to/from
$A$. References represent an oriented multi-graph on the union of nodes
of the callgraph and the varpool.

Finally, we implement a common infrastructure for jump functions.  Suppose that
an optimization pass see a function $A$ and it knows values of (some of) its
arguments.  The {\em jump function} \cite{IPACP, ipa-cp} describes the value of
a parameter of a given function call in function $A$ based on this knowledge
(when doing so is easily possible).  Jump functions are used by several
optimizations, such as the inter-procedural constant propagation pass and the
devirtualization pass.  The inliner also uses jump functions to perform
inlining of callbacks.

For easier development, the GCC pass manager differentiates between normal
inter-procedural passes and small inter-procedural passes.  An {\em small
inter-procedural pass} is a pass that does everything at once and thus it can
not be executed at the WPA time. It defines only the {\em
Execute} stage and during this stage it accesses and modifies the function
bodies.  Such passes are useful for optimization at LGEN or LTRANS time
and are used, for example, to implement early optimization before
writing object files. The simple inter-procedural passes can also be used for easier
prototyping and development of a new inter-procedural pass.

\subsection{Whole program assumptions, linker plugin and symbol visibilities}
\label{plugin}

Link-time optimization gives relatively minor benefits when used
alone.  The problem is that propagation of inter-procedural information does
not work well across functions and variables that are called or referenced by
other compilation units (such as from the dynamically linked library). We say
that such functions are variables are {\em externally visible}.

To make the situation even more difficult, many applications organize
themselves as a set of shared libraries, and the default ELF visibility rules
allow one to overwrite any externally visible symbol with a different symbol at
runtime.  This basically disables any optimizations across such functions and
variables, because the compiler cannot be sure that the function body it is
seeing is the same function body that will be used at runtime.  Any function
or variable not declared \code{static} in the sources degrades the quality of
inter-procedural optimization.

To avoid this problem the compiler must assume that it sees the whole program when doing
link-time optimization.  Strictly speaking, the whole program is rarely visible
even at link-time. Standard system libraries are usually linked
dynamically or not provided with the link-time information.  In GCC, the whole
program option (\code{-fwhole-program})  declares that every function and
variable defined in the current compilation unit\footnote{At link-time optimization the current unit
is the union of all objects compiled with LTO} is static, except for the
\code{main} function.  Since some functions and variables need to be referenced
externally, for example by an other DSO or from an assembler file, GCC also
provides the function and variable attribute \code{externally_visible} which can be
used to disable the effect of \code{-fwhole-program} on a specific symbol.

The whole program mode assumptions are slightly more complex in C++, where inline
functions in headers are put into COMDAT. {\em COMDAT} function and variables
can be defined by multiple object files and their bodies are unified at
link-time and dynamic link-time. COMDAT functions are changed to local only
when their address is not taken and thus un-sharing them with a library is not
harmful. COMDAT variables always remain externally visible, however for
readonly variables it is assumed that their initializers cannot be overwritten
by a different value.

The whole program mode assumptions do not fit well when shared
libraries are compiled with the link-time optimization.  The fact that ELF specification
allows overwriting symbols at runtime cause common problems with the increase of the dynamic
linking time and for this reason already common mechanisms to solve this
problem are available \cite{UrlichDSO}.

GCC provides the function and variable attribute {\tt visibility} that can be used to specify
the visibility of externally visible symbols (or alternatively an \code{-fdefault-visibility}
command line option).  ELF defines the \code{default}, \code{protected},
\code{hidden} and \code{internal} visibilities.  Most commonly used is
\code{hidden} visibility. It specifies that the symbol cannot be referenced from outside
of the current shared library. 

Sadly this information cannot be used directly by the link-time optimization in
the compiler since the whole shared library also might contain non-LTO objects
and those are not visible to the compiler. 

GCC solves this with the linker plugin.  The {\em linker plugin} \cite{plugin} is
an interface to the linker that allows an external program to claim the ownership of a given
object file.  The linker then performs the linking procedure by querying the plugin about
the symbol table of the claimed objects and once the linking decisions are complete, the plugin is allowed
to provide the final object file before the actual linking is made. The linker
plugin obtains the symbol resolution information which specifies which symbols
provided by the claimed objects are bound from the rest of a binary linked.

At the current time, the linker plugin  works only in combination with the Gold
linker,  but a GNU ld implementation is under development. 

GCC is designed to be independent of the rest of the tool-chain and aims to
support linkers without plugin support. For this reason it does not use the
linker plugin by default.  Instead the object files are examined before being
passed to the linker and objects found to have LTO sections are passed through
the link-time optimizer first.  This mode does not work for library
archives. The decision on what object files from the archive are needed
depends on the actual linking and thus GCC would have to implement the linker
by itself. The resolution information is missing too and thus GCC needs to make
an educated guess based on {\tt -fwhole-program}.  Without linker plugin
GCC also assume that symbols declared as {\tt hidden} and not referred by
non-LTO code by default.

The current behavior on object archives is suboptimal, since the LTO
information is silently ignored and LTO optimization is not done without any
report.  The user can then be easily disappointed by not seeing any
benefits at all from the LTO optimization.

The linker plugin is enabled via command line option \code{-fuse-linker-plugin}.
We hope that this becomes standard behavior in a near future. Many optimizations
are not possible without linker plugin support.

\section{Inter-procedural optimizations performed}
\label{ipasect}

GCC implements several inter-procedural optimization passes. In this section
we provide a quick overview and discuss their effectiveness on the test cases.

\subsection{Early optimization passes}
\label{earlyopts}
Before the actual inter-procedural optimization is performed, the functions are
early optimized.  Early optimization is a combination of the lowering passes
(where the SSA form is constructed), the scalar optimization passes and the
simple inter-procedural passes.  The functions are sorted in reverse postorder
(to make them topologically ordered for acyclic callgraphs) and all the passes
are executed sequentially on the individual functions. Functions are optimized
in this order and only after all passes have finished a given function is the
next function is processed.

Early optimization is performed at LGEN time to
reduce the abstraction penalty before the real inter-procedural
optimization is done.  Since this work is done at LGEN time
(rather than the link-time), it reduces the size of object files as well as the linking
time, because the work not re-done each time the object file is linked.  An object file 
is often compiled once and used many times. It is consequently beneficial to keep as
much of the work as possible in the LGEN rather than doing  link-time
optimization on unoptimized output from the front-end.

The following optimizations are performed:
\begin{itemize}
\item {\bf Early inlining}

Functions that have been already optimized earlier and
are very small are inlined into the current function.
Because the early inliner lacks any global knowledge of the program the
inlining decisions are driven by the code size growth, and only very small code
size growth is allowed.
\item {\bf Scalar optimization}

GCC currently performs  constant
propagation, copy propagation, dead code elimination, and scalar replacement.
\item {\bf Inter-procedural scalar replacement}

For static functions whose
address is not taken, dead arguments are eliminated and calling
conventions updated by promoting small arguments passed by reference
to arguments passed by value. Also when a whole aggregate is not needed, only the fields which are used
are passed, when this transformation is expected to simplify the resulting code \cite{IPA-sra}.

The main motivation for this pass is to make object-oriented programs easier to
analyze locally by avoiding need to pass the \code{this} pointer to simple methods.
\item {\bf Tail recursion elimination}
\item {\bf Exception handling optimizations}

This pass reduces the number of exception
handling regions in the program primarily by removing cleanup actions that were
proved to be empty, and regions that contains no code that which possibly throw.
\item {\bf Static profile estimation}

When profile feedback is not available, GCC attempts to guess the function profile
based on a set of simple heuristics  \cite{BallLarus,naspaer}.
Based on the profile, cold parts of the function body are identified (such
as parts reachable only from exception handling or leading to a function call
that never returns).  The static profile estimation can be controlled by user
via the \code{builtin_expect} builtin and the \code{cold} function attribute.
\item {\bf Function attributes discovery}

GCC has C and C++ language
extensions that allow the programmer to specify several function attributes as
optimization hints.  In this pass some of those attributes can be auto-detected.
In particular we detect functions that cannot throw
(\code{nothrow}), functions that never returns (\code{noreturn}), and
\code{const} and \code{pure} functions.  Const functions in GCC terminology
are functions that only return their value and their return value
depends only on the function arguments. For many optimization passes const functions behave like a
simple expression (allowing dead code removal, common subexpression
elimination etc.).  Pure functions are like const functions but are allowed to read global
memory. See \cite{GCCmanual} for details.
\item {\bf Function splitting pass}  This pass splits functions into headers
and tails to aid partial inlining.
\end{itemize}
Early optimization is very effective for reducing the abstraction penalty.  It
is essential for benchmarks that, for instance, use the Pooma library---early
inlining at the Tramp3d benchmark \cite{richi} causes an order of magnitude
improvement \cite{callgraph2007}.

The discovery of function attributes also controls the flow graph accuracy and
discovery of the noreturn functions significantly improves the effectivity of the static
profile estimation. Error handling and sanity checking code is  often
discovered as cold.

Early optimization is of lesser importance on code bases that do not have
significant abstraction or which are highly hand optimized, such as the Linux kernel. The 
drawback is that most of the work done by scalar optimizers needs to be re-done
again after inter-procedural optimization. Inlining, improved aliasing and other
facts derived from the whole program makes the scalar passes operate more effectively.
On such code bases early inlining leads to slowdowns in compile time while
yielding small benefits at best.

After early optimization unreachable functions and variables are removed.

\subsection{The whole program visibility pass}
This is the first optimization pass run at WPA when the callgraph of the whole
program is visible.  The pass decides which symbols are externally visible in
the current unit (entry point).  The decisions are rather tricky in details
and briefly described in Section~\ref{plugin}.  The pass also identifies
functions which are not externally visible and only called directly ({\em local
functions}). Local functions are later subject to more optimizations. For example,
on i386 they can use register passing conventions.

Unreachable functions and variables are removed from the program.  The removal of
unreachable functions is re-done after each pass that might render more
functions unreachable.
\subsection{IPA profile propagation ({\tt ipa-profile})}
Every function in GCC can be {\em hot} (when it is declared with the
\code{hot} attribute or when profile feedback is present), {\em normal}, {\em
executed once} (such as static constructors, destructors, \code{main} or
functions that never returns), or {\em unlikely executed}.

This information is then used to decide whether to optimize for speed or code
size. If optimization for size is not enabled, {\em hot} and {\em normal} functions are
optimized for speed (except for their cold regions), while functions {\em executed once} are optimized for speed only
inside loops. {\em Unlikely executed} functions are always optimized for size.

When profile feedback is not available, this pass attempts to promote the static
knowledge based on callers of the function.
\begin{itemize}
\item When all calls of a given function are unlikely (That is either the caller is {\em unlikely executed} or the function profile says that the
particular call is cold), the function is {\em unlikely executed}, too.
\item When all callers are {\em executed once} or {\em unlikely executed}, and call the function just once, the function is {\em executed once}
too. This is allows a number of executions of function {\em executed once} to be bound by known constant.
\item When all callers are executed only at startup, the function is also marked as executed only at startup.
This helps to optimize code layout of static constructors.
\end{itemize}

To optimize the program layout, the {\em hot} functions are placed in a separate
subsection of the text segment ({\tt .text.hot}). {\em Unlikely} functions are
placed in subsection {\tt .text.unlikely}. For GCC 4.6.0 we will also place
functions used only at startup into subsection {\tt .text.startup}.

While the pass provides an easy and cheap way to use the profile driven compilation
infrastructure to save some of code size, its benefits on large programs are
small. The decisions on what calls are cold are too conservative to give substantial
improvements on a large program.

For example, on Firefox, only slightly over 1000 functions are identified as cold,
accounting for fewer than than 1\% of functions in the whole program.  The pass
seems to yield substantial improvements only on small benchmarks, where code
size is not much of concern.

A more important effect of the pass is the identification of functions executed
at startup which we will discuss in Sections~\ref{ctororderings} and \ref{startuptime}.

\subsection{Constant propagation ({\tt ipa-cp})}
GCC's inter-procedural constant propagation pass \cite{ipa-cp} implements a standard algorithm using basic jump functions
\cite{IPACP}. Unlike the classical formulation, GCC pass does not implement return functions yet.
The pass also makes no attempt to propagate constants passed in static variables.

As an extension the pass also collects a list of types of objects passed to 
arguments to allow inter-procedural devirtualization when all types are known
and virtual method pointers in their respective virtual tables match.

Finally the pass performs cloning to allow propagation across functions which
are externally visible.  Cloning happens when all calls to a function are determined to pass the same
constant, but the function can be called externally too.
This makes the assumption that user forgot about \code{static} keyword and all the
calls actually come from the current compilation unit.  The original function
remains in the program as a fallback in case the external call happens.

More cloning would be possible: when a function is known to be used with two
different constant arguments, it would make sense to produce two clones;
this however does not fit the standard inter-procedural constant propagation
formulation and is planned for future function cloning pass.

A simple cost model is employed which estimates the code size effect of cloning.
Cloning which reduces overall program size (by reducing sizes of call sequences) is
always performed. Cloning that increase overall code size is performed only at
the \code{-O3} compilation level and is bound by the function size and overall unit
growth parameters.

The constant propagation is important for Fortran benchmarks, where it is often
possible to propagate arguments used to specify loop bounds and to enable
further optimization, such as auto-vectorization.  SPECfp2006 has several
benchmarks that benefit from this optimization.  The pass is also useful to propagate
symbolic constants, in particular \code{this} pointers when the method is only used on
single static instance of the object. Often various strings used for error
handling are also constant propagated.

So far relatively disappointing results are observed on the devirtualization
component of the pass. Firefox has many virtual calls and only about 200 calls
are devirtualized this way. Note that prior to this pass, devirtualization is
performed at local basis during the early optimizations.

\subsection{Constructor and destructor merging}
\label{ctororderings}


In this simple pass we collect all static constructors and
destructors of given priority and produce single function calling them
all that serves as a new static constructor or destructor.  Inlining will later
most likely produce a single function initializing the whole program.

This optimization was implemented after examining the disk access patterns
at startup of Firefox, see Section~\ref{startuptime}.  

\subsection{Inlining ({\tt ipa-inline})}
The inliner, unlike the early inliner, has information about the current unit
and profile, either statically estimated or read as profile feedback. As a
result it can make better global decisions than the early inliner.

The inliner is implemented as a pass which tries to do as much useful inlining as possible within
the bounds given by several parameters: the {\em unit growth} limits the code
size expansion on a whole compilation unit, {\em function growth} limits the expansion of
a single function (to avoid problems with non-linear algorithms in the
compiler), and {\em stack frame growth}.  Since the relative growth limits do not
work well for very small units, they apply only when the unit, function, or stack
frame is considered to be already large.  All these parameters are user-controllable
\cite{GCCmanual}.

The inliner performs several steps:
\begin{enumerate}
\item Functions marked with the \code{always_inline} attribute and all callers of functions
marked by the \code{flatten} attribute \cite{GCCmanual} are inlined.

\item Compilation unit size is computed.
\item Small functions are inlined.  Functions are considered small until
a specified bound on function body size is met.  The bound differs for functions
declared \code{inline} and functions that are auto-inlined. Unless \code{-O3}
or \code{-finline-functions} is in effect, auto-inlining is done only when
doing so is expected to reduce the code size.

This step is driven by a simple greedy algorithm which tries to inline functions in
order specified by estimated {\em badness} until the limits are hit.  After it
is decided that a given function should be inlined, the badness of its callers and
callees is recomputed.  The badness is computed as:
\begin{itemize}
 \item The estimated code size growth after inlining the function into all callers, when this growth is negative,
 \item The estimated code size growth divided by the number of calls, when profile feedback is available, or
 \item Otherwise computed by the formula
 $$c{\hbox{growth}\over \hbox{benefit}} + \hbox{growth for all}.$$
 Here {\em benefit} is an estimated
speedup of inlining the call, {\em growth} is the estimated code size growth caused by
inlining this particlar call, {\em growth for all} is the estimated growth caused by
inlining all calls of function, and $c$ is a sufficiently large magical constant.

The idea is to inline the most beneficial calls first, but also give some
importance to the information how hard it is to inline all calls of the given
function.
\end{itemize}

Calls marked as cold by the profile information are inlined only when doing so
is expected to reduce the overall code size.

At this step the inliner also performs inlining of recursive functions into
themselves.  For functions that do have large probability of (non-tail)
self-recursion this brings similar benefits as the loop unrolling.

\item Functions called once are inlined unless the function body or the stack frame growth limit is reached
or the function is not inlinable for an other reason.
\end{enumerate}

The inliner is the most important inter-procedural optimization pass and it is
traditionally difficult to tune.  The main challenge is to tune the inline
limits to get reasonable benefits for the code size growth, and to specify the
correct priorities for inlining. The requirements on inliner behavior depends
on particular coding style and the type of application being compiled.

 GCC has a relatively simple cost metric compared to
other compilers \cite{HP, Zhao}.  Some other compilers attempt to estimate, for
example, the effect on the instruction cache pollution and other parameters, combining
them into a single badness value.  We believe that these ideas have serious
problems in handing programs with a large abstraction penalty. The code seen by the
inliner is very different from the final code and thus it is very difficult to
get reasonable estimates.  For example, the Tramp3d benchmark has over 200
function calls in the program before the inlining for every operation performed
at execution time by the optimized binary.  As a result, inline heuristics have
serious garbage-in garbage-out problems.

We try to limit the number of metrics we use for inlining in GCC.  At the
moment we use only code size growth and time estimates. We employ several
heuristics predicting what code will be optimized out, and plan to extend them
more in the future. For example, we could infer that functions optimize better
when some of their operands are a known constant. We combine early inlining to
reduce the abstraction penalty with careful estimates of the overall code size
growth and dynamic updating of priorities in the queue. GCC takes into account
when offline copies of the function will be eliminated.   Dynamic updating of
the queue improves the inliner's ability to solve more complex scenarios over
algorithms processing functions in a pre-defined order. On the other hand this
is a source of scalability problems as the number of callers of a given
function can be very large.  The badness computation and priority queue
maintenance has to be effective.  The GCC inliner seems to perform well when
compared with implementations in other compilers especially on benchmarks with
a large C++ abstraction penalty.

\subsection{Function attributes ({\tt ipa-pure-const})}

This is equivalent to the pass described in Section~\ref{earlyopts} but
propagates across the callgraph and is thus able to propagate across boundaries
of the original source files as well as handle non-trivial recursion.

\subsection{MOD/REF analysis ({\tt ipa-reference})}

This pass \cite{ipa-reference} first identifies static variables which are never written to as
read-only and static variables whose address is never taken by a simple
analysis of the IPA reference information.

For static variables that do not have their address taken ({\em non-escaping variables}),
the pass then collects information on which functions read or modify them.
This is done by simple propagation across the callgraph, with strongly connected
regions being reduced and final information is stored into bitmaps which are later used by
the alias analysis oracle.

To improve the quality of the information collected, a new function attribute \code{leaf}
was introduced.  This attribute specifies that the call to an external {\tt leaf} function may return
to current module only by returning or with exception handling. As a result the calls
to \code{leaf} functions can be considered as not accessing any
of the non-escaping variables.

This pass has several limitations: First, the bitmaps tends to be quadratic
and should be replaced by a different data structure. There are also
a number of possible extensions for the granularity of the information collected. The
pass should not work on the level of variables, but instead analyze fields of structures independently.
Also the pass should not give up when the address of variable is passed to e.g.
a {\tt memset} call.

It is expected that the pass will be replaced by the inter-procedural points-to
analysis once it matures.

MOD/REF is effective for some Fortran benchmarks. On programs written in a
modern paradigm it suffers from the lack of static variables initialized.  The
code quality effect on Firefox and GCC is minimal. Enabling the optimization
save about 0.1\% of GCC binary size.

\subsection{Function reordering ({\tt ipa-reorder})}
\label{ipa-reorder}

This is an experimental pass we implemented while analyzing Firefox startup
problems.  It specifies the order of the functions in the final binary by concatenating
the callgraph functions in  priority order, where the priority is given by the likeliness
that one function will call an other.  This pass increase code locality and
thus reduces the number of pages that needs to be read at program startup.

During typical Firefox startup, poor code locality causes 84\% of the
\code{.text} section to be paged in by the Linux kernel while only 19\% is
actually needed for
program execution. Thus there is room for up to a $3\times$ improvement in library
loading speed and memory usage \cite{Glandiumblog}.
At the time of writing the paper we cannot demonstrate consistent improvements
in Firefox.  When starting the GCC binary, the operating system needs to read
about 3\% fewer pages.  It is not decided yet if the pass will be included in GCC~4.6.0
release.

\subsection{Other experimental passes}

GCC also implements several optimization passes that are not yet ready for
compiling larger application. In particular inter-procedural points to
analysis, a structure reordering pass \cite{structreorg} and a matrix
reorganization pass \cite{maxtrixreorg}. 

\subsection{Late local and inter-procedural optimization}

At the LTRANS stage the functions of a given callgraph partition are compiled in the
reverse postorder of the callgraph (unless the function reordering pass is
enabled).  This order allows GCC to pass down certain information from callees to
callers.  In particular we re-do function attribute discovery and propagate
the stack frame alignment information.  This reduces the resulting size
of the binary by additional 1\% (on both Firefox and GCC binaries). 

\section{Compiling large applications}
\label{largesect}

In this section we describe major problems observed while building large
applications.  We discuss the performance of the GCC LTO implementation and its
effects on the application compiled.

Ease of use is a critical consideration for compiler features.  For example,
although compiling with profile feedback can yield large performance
improvements to many applications \cite{naspaer} it is used by few software
packages.  Even programs which might easily made to benefit from profile-guided
optimizations, such as scripting language interpreters, have not widely adopted
this feature. One of the main design goals of the LTO infrastructure was to
integrate as easily as possible into existing build setups \cite{LTOproposal,
WHOPR}.  This has been partially met.  In many cases it is enough to add
\code{-flto -fwhole-program} as a command line option.  

In more complex packages, however, the user still needs to understand the use of
\code{-fuse-linker-plugin} (to enable the linker plugin to support object
archives and better optimization) as well as \code{-fwhole-program} to enable
the whole program assumptions.  GCC 4.6.0 will introduce the WHOPR mode (enabled via
\code{-flto}) command line option. The user then needs to specify the parallelism
via \code{-flto=}$n$.  We plan to add auto detection of GNU Make that allows
parallel compilation via its job server (controlled by well known \code{-j}
command line option).  The job server can be detected using a environment
variable, but it requires the user to add \code{+} at the beginning of the \code{Makefile} rule.

In this section we concentrate mostly on our experiences building Firefox
and the GCC itself with link-time optimization.

Firefox is a complex application. It consist of dozens of libraries totaling
about 6 millions of lines of C and C++ code.  In addition to being a large
application it is used heavily by desktop users.

Many libraries built as part of Firefox are developed by third parties with
independent coding standards.  As such they stress areas of the link-time
optimization infrastructure not used at all by portable C and C++ programs
such as the ones present in the SPEC2006 benchmark suite.  For this reason we chose Firefox
as a good test for the quality and practical usability of GCC LTO support.  We
believe that by fixing numerous issues arising during Firefox build we also
enabled GCC to build many other large applications.

On the other hand the the GCC compiler itself is a portable C application.
The implementation of its main module, the compiler binary itself, consist of about
800~000~lines of hand written C code and about 500~000~lines of code
auto-generated from the machine description. We test the effect of the link-time
optimization on GCC itself especially because the second author is very
familiar with the code base. 

At the time of writing this paper, both GCC and Firefox compile and work with
LTO.  Firefox requires minor updates to the source code---In particular, we had
to annotate the variables and functions used by {\tt asm} statements.  This is
done with attribute \code{used} \cite{GCCmanual}.

\subsection{Compilation times}

Compilation time increases are always noticeable when switching from the normal
compilation to the link-time optimizing environment.  Linking is more difficult to
distribute and parallelize than the compilation itself. Moreover, during
development, the program is re-linked many times after modifications in some of
source files.  In the file-by-file compilation mode only modified files are re-compiled
and re-linking is relatively fast.  With LTO most of the optimization work is lost and
all optimizations at link-time have to be redone again.

With the current GCC implementation of LTO, the overall build time is expected to
double at least. This is because of the use of ``fat'' object files.  This
problem is will be solved soon by introduction of ``slim'' object files.

The actual expense of streaming IL to object files is minor during the compilation stage, so we
focus on actual link-times. We use an 24~core AMD workstation for our testing.
The ``fat'' object files are about 70\% larger than object files which contain assembler
code only. (GCC use zlib to compress the LTO sections).

\subsubsection{GCC}

Linking GCC in single CPU LTO mode needs 6~minutes and 31~seconds. This is similar to the time
needed for the whole non-LTO compilation (8~minutes and 12~seconds). Consequently
time the needed to build the main GCC binary from the scratch is about 15~minutes.

The overall increase of build time of GCC package is bigger. Several compiler
binaries are built during the process and they all are linked with a common
backend library.  With the link-time optimizations the backend library is thus
re-optimized several times.  We do not count this and instead measure time
needed to build only one compiler binary.

The most time-consuming steps of the link-time compilation are the the following:
\begin{itemize}
\item Reading the intermediate language from the object file into the compiler: 3\% of the overall compilation time.
\item Merging of declarations: 1\%.
\item Outputting of the assembly file: 2\%.
\item Debug information generation (\code{var-tracking} and \code{symout}): 8\%.
\item Garbage collection: 2\%.
\item Local optimizations consume the majority of the compilation time.

 The most expensive components are:
 partial redundancy elimination (5\%), GIMPLE to RTL expansion (8\%), RTL level dataflow analysis (11\%), instruction combining (3\%),
register allocation (6\%), scheduling (5\%).
\end{itemize}
The actual inter-procedural optimizations are very fast, with the slowest being the inline
heuristics. It still accounts for less than 1\% of the compilation time (1.5
seconds).

In WHOPR mode with the use of all 24~cores of our testing workstation
we reduce the link time to 48~seconds. This is also faster than the
parallelized non-LTO compilation, which needs 56~seconds.  As a result, even in
a parallel build setup, the LTO accounts for a two-fold slowdown.  The slower
non-LTO build time is partly caused by the existence of large, auto-generated
source files, such as {\tt insn-attrtab.c}, which reduce the overall parallelism.  WHOPR
linking has the advantage of partitioning the {\tt insn-attrtab.c} into multiple
pieces.

The serial WPA stage takes 19~seconds. The most expensive steps are:
\begin{itemize}
\item Reading global declarations and types: 28\% of the overall time taken by the WPA stage.
\item Merging declarations: 6\%.
\item Inter-procedural optimization: 9\%. 
\item Streaming of object files to be passed to LTRANS: 42\%.
\end{itemize}
The rest of the compilation process, including all parallel LTRANS stages, and the actual linking consume 
29~seconds.

\subsubsection{Firefox}

Linking Firefox in single CPU LTO mode needs 19~minutes and 29~seconds (compared to
39~minutes needed to build Firefox from scratch in non-LTO mode).
The time is distributed as follows:
\begin{itemize}
\item Reading of the intermediate language from the object file into the compiler: 7\%.
\item Merging of declarations: 4\%.
\item Output of the assembly file: 3\%.
\item Debug information generation is disabled in our builds.
\item Garbage collection: 2\%.
\item Local optimizations consume majority of the compilation time.

 The most expensive components are:
 operand scan (5\%), partial redundancy elimination (5\%), GIMPLE to RTL expansion (13\%), RTL level dataflow analysis (5\%), instruction combining (3\%),
register allocation (9\%), scheduling (3\%).
\end{itemize}

The WHOPR mode reduces the overall link-time to 5~minutes and 30~seconds. WPA stage takes
4~minutes 24~seconds. This compares favorably to non-LTO parallel compilation
which take 9~minutes and 38~seconds. The source code of Firefox is organized into
multiple directories leading to less parallelism exposed to Make.
The most expensive steps are:
\begin{itemize}
\item Reading global declarations and types: 24\%.
\item Merging declarations: 20\%.
\item Inter-procedural optimization: 8\%. 
\item Streaming of object files to be passed to LTRANS: 28\%.
\item Callgraph and WPA overhead (callgraph merging and partitioning): 12\%.
\end{itemize}

The fact that the link-time optimization seems to scale linearly and maintain
a two-fold slowdown can be seen as a success. WHOPR mode
successfully enables GCC to use parallelism, to noticeably reduce the build time.  It
is however obvious that the intermediate language input and output is a
bottleneck.  We can address this problem in two ways.  First, we can reducing
the number of global types streamed by separating debug information and analyzing the
reason why so many types and declarations are needed.  Second, we can optimize the
on-disk representation.  By solving these problems, the WPA stage can become 
several times faster, since the actual inter-procedural optimization seems to
scale very well.  Just shortly before finishing the paper, Richard G\"unther 
submitted a first patch to reduce the number of declarations at WPA stage to about 1/4th.
This demonstrates that there is quite a lot of space for improvement left here.

Because WHOPR makes linking faster than the time needed to build non-LTO
application, there is hope that with the introduction of ``slim'' objects, the
LTO build times will be actually shorter than non-LTO for many applications.
This is because slow code generation is better distributed to multiple CPUs
with WHOPR than with avreage parallel build machinery.

Linking time comparable to the time needed to rebuild the whole
application is still very negative for the edit/recompile/link experience of
the developers. It is not expected that developers will use LTO optimization at all
stages of development, but still optimizing for quick re-linking after a
local modification is important.  We plan to address this by the introduction of an
incremental WHOPR mode, as discussed in Section~\ref{Conclussion}.
\subsection{Compile time memory usage}

A traditional problem in GCC is the memory usage of its intermediate
languages.  Both GIMPLE and RTL require an order of magnitude more memory than 
the size of the final binary. While the intermediate languages needs to keep more
information than the actual machine code, other compilers achieves this with a lot slimmer
ILs \cite{LLVM}.
In addition to several projects to reduce memory usage of GCC (see, for example,
\cite{GIMPLE}), this was also one of motivations for designing WHOPR to avoid the
 need to load the whole program into the memory at once.

In LTO mode memory usage peaks at 2GB for the GCC compilation and 8.5GB on the Firefox
compilation.

In WHOPR mode, the WPA stage operates only at the callgraph and optimization
summaries, while the LTRANS stage sees only parts of the program at a time. In
our tests we configure WHOPR to split the program into 32~partitions.  
Compiling large programs is currently dominated by the memory usage of the WPA
stage: in a 64-bit environment the memory usage of the compilation of the GCC
binary peaks at 415MB, while the compilation of Firefox peaks slightly
over 4GB.  The actual LTRANS compilations do not consume more than 400MB,
averaging 120MB for the GCC compilation and 300MB for Firefox. 

The main binary of the GCC compiler (cc1) is 10MB, so the GCC compile-time
memory consumption is still about 50~times larger than the size of the program.
Compiling Firefox uses about 130~times more memory than the size of the
resulting binary.  This is a serious problem especially for a 32-bit
environment, where the memory usage of a single process is bound by the address
space size.  In 32-bit mode, GCC barely fits in the address space when
compiling Firefox!

The main sources of the problem are the following:
\begin{itemize}
\item GCC is mapping the source object files into memory.  This is fast, but
careless about address space limits in 32bit environments.  For GCC this accounts to about 170B of
the address space. 

It is not effective to open one file at a time, since the files are read in
several stages, each stage accessing all files. Clearly a more effective
scheme is still possible.
\item A large amount of memory is occupied by the representation of types and declarations.
Many of these are not really needed at the WPA stage and should be streamed independently.
For GCC this accounts for 260MB of the memory.
\item The MOD/REF pass has a tendency to create overly large bitmaps.  This is not
problem when building GCC or Firefox, but it can be observed on some benchmarks in
the SPEC2006 test-suite, where over 100MB of bitmaps are needed.
\end{itemize}

Note that a relatively small amount (about 52MB in compilation of GCC) is used by the actual
callgraph, varpool, and other data structures used for the inter-procedural
optimization.

The memory distribution of the compilation of Firefox is very similar to one seen at GCC compilation. The
percentage of memory used by declarations and types is even higher --- about
3.7GB. This is because C++ language implies more types and longer identifiers.

Similarly to the compilation time analysis, we can identify declarations and types
as being a major problem for scalability of GCC LTO.  

\subsection{Code size and quality}

Link time optimization promises both performance improvements as well as code size
reductions.  It is not difficult to demonstrate benchmarks where cross module
inlining and constant propagation cause substantial performance improvements.
However in applications that has been profiled and hand optimized for a single-file
compilation model (this is the case of both Firefox and GCC), the actual
performance improvements are limited. This is because the authors of the
software already did by hand most of the work the inter-procedural optimizer
would otherwise do.

In the short term, we expect the value of link-time optimizations on such
applications to be primarily in the reduction of the code size.  It is harder
use hand optimization to reduce the overall size of the application than to
increase performance. Most program spends most of their time in a rather small
portion of the code, so one can optimize for speed only the hot code.  But to
decrease overall program size, one must tune the whole application.

Once more widely adopted, the link-time optimization will simplify the task of
developers by reducing the amount of hand-tuning needed.  For example, with LTO
it is not necessary to place short functions into headers for better inlining.
This allows a cleaner cut between interface and implementation.

\subsubsection{Firefox}

When compiled with link-time optimization (using the \code{-O3} optimization
level) the size of Firefox main module reduce from 33.5MB to
31.5MB, a reduction of 6\%. 

Runtime performance tests comparing Firefox built without LTO to
Firefox built with LTO using the \code{-O3} optimization level are shown
in the Figure~\ref{Firefox}.
\begin{figure}[t]
\begin{center}
\begin{tabular}{|l|r|l|}
\cline{1-2}
benchmark name & speedup \\
\cline{1-2}
dromeao css & 1.83\% \\
tdhtml & -0.54\% \\
tp\_dist & 0.50\% \\
tsvg & 0.07\% \\
\cline{1-2}
\end{tabular}
\end{center}
  \caption{Firefox performance.}
\label{Firefox}
\end{figure}

When optimizing for size (\code{-Os}),  early reports at building Firefox  with
the LLVM compiler and LTO enabled claim to save 13\% of the code size
compared to GCC non-LTO build with the same settings \cite{Rafael}. Comparing the GCC
non-LTO build (28.2MD) with the GCC LTO build (25.3MB) at
\code{-Os}, we get a 11\% smaller binary.  

We also observed that further reductions of the code size are possible by
limiting the overall unit growth (\code{--param inline-unit-growth} parameter).
We found, that limiting overall growth to 5\% seems to give considerable
code size saving (additional 12\%), while keeping most of the performance benefits of
\code{-O3}. Since this generally applies to other big compilation units too, we
plan to re-tune the inliner to automatically cut the code size growth with 
an increasing unit size.

The non-LTO \code{-O3} build is 18\% bigger than the non-LTO \code{-Os} build.
Consequently enabling LTO (and tweaking the overall program growth) has a code
size effect comparable to switching from the aggressive optimization for speed
to the aggressive optimization for size.  LTO however has positive performance
effects, while \code{-Os} is reported to be about 10\%--17\% slower.

\subsubsection{GCC}

GCC by default uses the \code{-O2} optimization level to build itself.  When
compiled with the link-time optimization, the GCC binary shrinks from
10MB to 9.3MB, a reduction of 7\%.  The actual speedups are
small (within noise level) because during the work on the link-time optimization we
carefully examined possibilities for cross-module inlining and reorganized
the sources make them possible in single-file compilation mode, too.

Some improvements are seen when GCC is compiled with \code{-O3}. The non-optimizing
compilation of C programs is then 4\% faster.  The binary size is 11MB.

\subsubsection{SPEC2006}

For reference we include SPEC2006 results on an AMD64 machine, comparing
the options:\\
 \code{-O3 -fpeel-loops -ffast-math}\\
\code{-march=native}\\ with the same options plus\\
\code{-flto -fwhole-program}.

In Figures~\ref{specint} and \ref{specfp} the first column compares SPEC rates
(bigger is better), and the second column compares executable sizes (smaller is
better).

\begin{figure}[t]
\begin{center}
\begin{tabular}{|l|r|r|}
\cline{1-3}
	  & speedup & size \\
\cline{1-3}
perlbench &+1.4\%   &  +4\%\\
bzip2     &+2.6\%   &  -45\%\\
gcc       &-0.3\%   &  +1.2\%\\
mcf       &+1.9\%   &  -33\%\\
gobmk     &+3.4\%   &  +1.8\%\\
hmmer     &+0.8\%   &  -55\%\\
sjeng     &+1.2\%   &  -11\%\\
libquantum&-0.5\%   &  -61\%\\
h264ref   &+7.0\%   &  -9\%\\
omnetpp   &-0.8\%   &  -11\%\\
astar     &-1.3\%   &  -20\%\\
\cline{1-3}
\end{tabular}
\end{center}
  \caption{SPECint 2006.}
\label{specint}
\end{figure}
\begin{figure}[t]
\begin{center}
\begin{tabular}{|l|r|r|}
\cline{1-3}
	  & speedup & size \\
\cline{1-3}
bwaves    &0\%  (+15\%)&  -27\%\\
gamess    &-0.7\%      & -50\%\\
milc      &+2.2\%      & -26\%\\
zeusmp    &+0.4\%      & -27\%\\
gromacs   &0\%	   &-18\%\\
cactusADM &-0.8\%      & -42\%\\
leslie3d  &-2.1\%  (0\%)&  +0.6\%\\
namd      &0\%         & -40\%\\
soplex    &+1.5\%      & -50\%\\
povray    &+5\%        & -2.3\%\\
calculix  &1.1\%	   &-38\%\\
GemsFDTD  &0\%         & -70\%\\
tonto     &-0.2\%      & -25\%\\
lbm       &+3.2\%      & 0\%\\
wrf       &0\%	   &-36\%\\
sphinx3   &+2.9\%      & -32\%\\
\cline{1-3}
\end{tabular}
\end{center}
  \caption{SPECfp 2006.}
\label{specfp}
\end{figure}
The results show significant code size savings derived from improved
inlining decisions and the whole program assumption (especially in code bases that
do not use the \code{static} keyword in declarations consistently).  The
performance is also generally improved.  The  Bwaves and leslie3d benchmarks
demonstrate a problem in the GCC static profile estimation algorithm where
inlining too many loops together causes a hot part of program to be predicted
cold.  The results in parenthesis shows the results with hot/cold decisions
disabled.

We did not analyze the regressions in astar or zeusmp yet. We also excluded the
xalancbmk and dealII benchmarks, since they do not work with the current GCC
build\footnote{Both benchmarks were working with GCC LTO in the past.}.

\subsection{Startup time problems}
\label{startuptime}
One of Firefox's goals is to start quickly. We devote a special section to this
problem, because it is often overlooked by tool-chain developers. 
Startup time issues are not commonly visible in common benchmarks, which generally
consist of small- to medium-sized applications.

Unfortunately, it turns out that currently Linux + GNU tool-chain is ill-suited
for starting large applications efficiently. Many of the other open source
programs of Firefox's size (e.g. OpenOffice, Chromium, Evolution) suffer
from various degrees of slow startup.

There are various ways to measure startup speed. In this paper we will focus on
cold startup as a worst-case scenario.

\subsubsection{Overview of Firefox startup}

The {\tt firefox-bin} ``stub'' calls into \code{libxul.so}, which is a large library that
implements most of Firefox's functionality. For historical reasons the rest of
Firefox is broken up into 14 other smaller libraries (e.g. the {\tt nspr} portability
library, {\tt nss} security libraries). Additionally, Firefox depends on a large
number of X/GTK/GNOME libraries.

{\bf Components of Firefox startup}

The Firefox startup can be categorized into the following phases:

\begin{enumerate}
\item Kernel loads the executable.
\item The dynamic library loader then loads all of the prerequisite libraries.
\item For every library loaded, initializers (static constructors) are run.
\item Firefox \code{main(}) and the rest of application code is run.
\item The dynamic library loader loads additional libraries.
\item Finally, a browser window is shown.
\end{enumerate}

It may come as a surprise that steps 2--3 currently dominate Firefox
startup on Linux. This being a cold startup, IO dominates. There are 2 kinds of
IO, explicit IO done via explicit {\tt read()} calls and implicit IO facilitated by
\code{mmap()}.

Most of the overhead comes from not using \code{mmap()} carefully. This IO is
triggered by page faults which are essentially random IO. Typically a page fault
causes 128KB of IO around the faulted page. For example, it takes 162
page faults (20MB/128KB) to page in the {\tt .text} section for {\tt libxul.so}, Firefox's main
library. Each page fault incurs a seek followed by a read. Hard drive
manufactures specify disk seeks ranging from 5ms to 14ms(7200 to 4200)\footnote{SSDs do not suffer from disk seek latency. However, there are still IO delays ranging from 0.1ms to 2s depending on the types of flash and controllers used \cite{Anandtech}.}. In practice a
missed seek seems to cost 20-50ms.

Modern storage media excels at bulky IO, but random IO in small chunks keeps
devices from performing at their best.

\begin{figure}[t]
\begin{center}
\includegraphics[width=8cm]{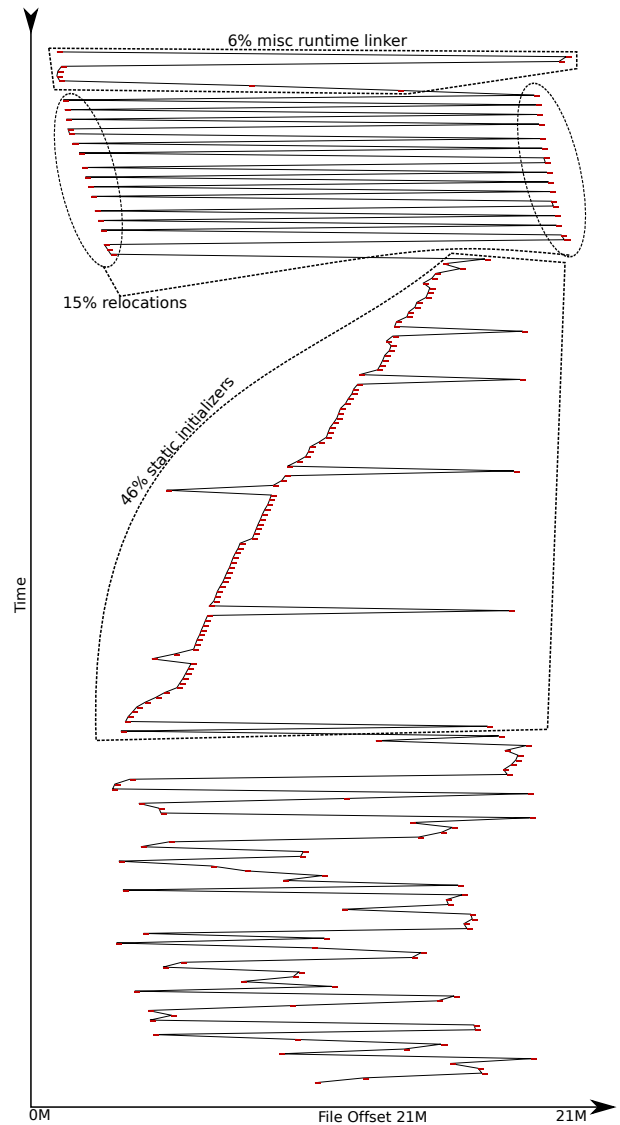}
\end{center}
  \caption{page faults that occur while loading {\tt libxul.so} from disk.}
\label{accesspattern}
\end{figure}
Figure~\ref{accesspattern} illustrates the page faults which occur while loading {\tt libxul.so} from
disk.

We now examine each of the stages of loading Firefox in more detail:

{\bf Loading {\tt firefox-bin}}

This is cheap because {\tt firefox-bin} is a small executable, weighing in at only 48K. This is smaller
than Linux's readahead, so loading {\tt firefox-bin} only requires a single read from
disk. The dynamic loader then proceeds to load various libraries that
{\tt firefox-bin} depends on.

{\bf {Dynamic linker ({\tt ld.so})}}

In the shared-lib-happy Linux world, the runtime linker, {\tt ld.so}, loads a total of
105 libraries on {\tt firefox-bin}'s behalf. It is essential that {\tt ld.so}
carefully avoids performing unnecessary page faults and that the compile-time
linker arranges the binary to facilitate that.

{\tt ld.so} loads dependent libraries, sets up the memory mappings,
zeroes .bss\footnote{.bss follows .data which usually does not end on a page
boundary.}, etc. As seen in Figure~\ref{accesspattern}, relocations are the
most expensive part of library loading. Prelink addresses some of the
relocation overhead on distributions like Fedora.

{\bf Static Initializers}

Once a library is loaded by {\tt ld.so}, the static initializers for that library are
enumerated from the {\tt .ctors} section and executed.

In Firefox static initializers arise from initializing C++ globals with
non-primitives. Most of time they are unintentional, and can even be caused by
including standard C++ headers (e.g. \code{iostream}). Better inlining of simple
constructors into POD-initialization could alleviate this problem.

Compounding the problem of unintentional static initializers is the fact that GCC
treats them inefficiently. GCC creates a static initializer function
and a corresponding {\tt .ctors} entry for every compilation unit with static
initializers. When the linker concatenates the resulting object files, this
has the effect of evenly spreading the static initializers across the entire
executable.

The GCC runtime then reads the {\tt .ctors} entrys and executes them in reverse
order. The order is reverse to ensure that any statically linked libraries are
initialized before code that depends on them.

The combined problem of the abundance of C++ static initializers, their layout
in the program, and the lack of special treatment by the linker means that
executing the {\tt .ctors} section of a large C++ program likely causes its
executable to be paged in backwards!

We noticed that, for example, the Microsoft C++ compiler and linker group static
initializers together to avoid this problem.

{\bf Application Execution}

A nice property of static initializers is that the order of their execution is
known at compile time. Once the application code starts executing, {\tt .text} is
paged in in an essentially random pattern.  During file-by-file compilation, functions are
laid out based on the source file implementing them, with no relation to their
callgraph relationships. This results in poor page cache locality.

\subsubsection{Startup time improvements within reach of the LTO infrastructure}

While working on enabling LTO compilation of Firefox, we also experimented with
several simple optimizations targeted to improve the startup time.  The most obvious
transformation is to merge the static initializers into a single function for
better code locality. This is implemented as a special constructor merging pass,
see Section~\ref{ctororderings}.

This transformation almost eliminates the part of disk access graph attributed
to execution of static initializers. Ironically for Firefox itself this only delays most
of disk accesses to a later stage of the Firefox startup because a lot of code is
needed for the rest of startup process. Other C++ applications however suffer
from the same problem. If the application does less work during the rest of
startup, its startup time will benefit noticeably.

To further improve the code locality of the startup procedure, we implemented
a function reordering pass, see Section~\ref{ipa-reorder}. Unfortunately we
were not yet able to show consistent improvements of this pass on  Firefox
itself.  The problem is that compiler, besides the execution of static initializers, has
little information about rest of startup process and just grouping function
based on their relative references seems not to interact well with kernel's
readahead strategy.

To improve the static analysis, further work will be needed to track virtual
calls.  The design of the Firefox APIs allows a lot of devirtualization which GCC is not
currently capable of. When devirtualization fails we can produce speculative callgraph
edges from each virtual call to every virtual method of a compatible type (may
edges) and use them as hints for ordering.  Clearly may edges are one
of main missing parts of the GCC inter-procedural optimization infrastructure.  It
is however not clear how much potential benefit these methods will have in practice, as
the actual problem of lacking knowledge of the startup procedure
remains. Improving GCC devirtualization capabilities alone would be however
important improvements in its own.

Locality is also improved by aggressive inlining of functions which are called once.

In the near future we plan to experiment more with the profile feedback
directed optimization in combination with LTO.  With profile feedback
available, the actual problem of ordering functions for better startup time is
a lot easier: All we need is to extend the program instrumentation to record
the time when a given function was invoked first and order functions according
to their invocation times.

Finally data layout can be optimized.  Data structures with static constructors
can be placed at a common location to reduce the amount of paging as well.

\section{Conclusion}
\label{Conclussion}
The link-time optimization infrastructure in GCC is mature enough to compile
large real-world applications.  The code quality improvements are comparable
with other compilers.  

Unlike heavily benchmark-optimized compilers, GCC usually produces smaller
binaries when compiling with LTO support than without.  We expect that the code size
effect will be one of main selling points of LTO support in GCC in the near
future. Code size and  code locality are both very important factors
affecting the performance of large, real-world applications.  The
link-time optimization model allows the compiler to make substantial
improvements in this area. The size of the Firefox binary built with link time
optimization for speed is comparable to the size the Firefox binary built
with file-by-file optimization for size.

LTO also brings runtime performance improvements.  The magnitude of these
improvements largely depends on how much the benchmarked application was
profiled and hand-tuned for the single-file compilation model.
Many of the major software packages today went through this tuning.  Consequently
the immediate benefits of classical inter-procedural optimizations are limited.
Both GCC and Firefox show a runtime improvement of less than 1\% with LTO.

GCC also exposes a lack of tuning for the new environment as seen in two
larger regressions in SPECfp2006 benchmark.  We hope to re-tune the inliner and
profile estimation before GCC 4.6.0 is released.  More benefits are possible
when the compiler enables some of its more aggressive optimizations (such as
code size expanding auto-inlining) by default.   This can be done because the compiler is significantly
more aware of the global tradeoffs at link time than at compile time.  Enabling automatic inlining with a small
overall program growth limit (such as 5\%) improves GCC performance by 4\%.
This suggests that in the future GCC should enable some of those code expanding
optimizations by default during link-time optimization, with the overall code size
growth bounded to reasonable settings.

GCC also lacks some of more advanced inter-procedural optimizations available in other compilers.  To make
our inter-procedural optimizer complete, we should introduce more
aggressive devirtualization, points-to analysis, a function specialization pass
and, a pass merging functions with identical bodies. The callgraph module also
lacks support for may edges representing possible targets of indirect calls.
There is also a lot of potential in implementing data structure layout
optimizations \cite{Layout}, such as structure field reordering for better
locality.  Some of these passes already exist in the form of experimental
passes \cite{structreorg, maxtrixreorg}. Getting these passes into production
quality will involve a lot of work.

More performance improvements are possible by a combination of the link
time optimizations and profile feedback directed optimizations \cite{LIPO}: link-time
optimization gives the compiler more freedom, while profile feedback tells
the compiler more precisely which transformations are beneficial.  Immediate benefits
can be observed, for example, in the quality of inlining decisions, code layout,
speculative devirtualization and code size.  GCC has
profile feedback support \cite{naspaer} and it is tested to work with the link
time optimization.  More detailed study of the benefits is however out of the
scope of this paper.

There are a number of remaining problems.  The on-disk representation of the
intermediate language is not standardized at all. This implies  that all files
needs to be recompiled when the compiler or compiler options change. This limits
possibilities of distributing LTO-compiled libraries.  The memory usage is 
comparable with other compilers (such as MSVC), yet a number of improvements are
possible, especially by reducing the number of declarations and types streamed.

A GCC-specific feature, the WHOPR mode, allows parallel compilation. Unlike a
multi-threaded compilation model, it allows distributed compilation, too,
although it is questionable how valuable this is, since today and tomorrow's
workstation machines will likely have a good deal of parallelism available
locally.  Increasing the number of parallel compilations past the 24 we used in
our testing would probably have few benefits, since the serial WPA stage would
likely dominate the compilation.

The WHOPR mode makes a clean cut in-between inter-procedural propagation and local
compilation using optimization summaries.  This invites the implementation
of an incremental mode, where during re-compilation the whole work is not
re-done. Only the inter-procedural passes would be re-run and the assembly code of
functions whose body nor summary changed would reused from the previous run.
This is planned for the next GCC releases.

Still, before GCC~4.6.0 is released (and probably even for GCC~4.7.0) there is
a lot of work left to do on correctness and feature completeness of the basic
link-time optimization infrastructure.  The main areas lacking include
debugging information, which is a lot worse than in file-by-file compilation.
At the time of writing this paper enabling debug information also leads to
compiler crash when building Firefox.  Clearly this is important problem as no
major project will use a compiler that does not produce usable debug
information for building of official binaries.

\section{Acknowledgment}
We thank to Andi Kleen and Justin Lebar for several remarks and corrections
which significantly improved the quality of this paper.


\end{document}